\documentclass[11pt]{article}
\usepackage{epsfig}
\usepackage{cite}

\newlength{\pictwidth}
\begin{document}

\setcounter{page}{1}

\begin{center}
\begin{Large}
{Spin Physics with STAR at RHIC}
\end{Large}

\vspace*{0.2in}
J. Kiryluk  for the STAR Collaboration \\
\vspace*{0.1cm}
{\footnotesize{\em{Massachusetts Institute of Technology \\ 77 Massachusetts Ave., Cambridge MA  02139-4307, USA\\E-mail: joanna@lns.mit.edu}}} \\

\end{center}

\vspace*{0.2in}

\begin{abstract}
STAR collected data in proton-proton collisions at $\sqrt{s}=200$ GeV with transverse and 
longitudinal beam polarizations during the initial running periods in 2002--2004 at 
the Relativistic Heavy Ion Collider at Brookhaven National Laboratory.  
Results on the single transverse spin asymmetries in the production of high energy forward neutral pions 
and of forward charged hadrons will be presented.
Data have been obtained for double longitudinal asymmetries in inclusive jet production in 2003 and 2004.
These data provide sensitivity to the polarization of gluons in the proton.
In the future, we aim to determine the gluon polarization over a wide kinematic range using coincidences 
of direct photons and jets. Furthermore, we aim to determine the polarizations of the $u$, $\bar{u}$, 
$d$ and $\bar{d}$ quarks in the proton by measuring single longitudinal spin asymmetries in the production 
of weak bosons at $\sqrt{s} = 500$ GeV.
\end{abstract}

\vspace*{0.2in}

STAR is one of two large experimental facilities at the Relativistic Heavy Ion Collider (RHIC) at 
Brookhaven National Laboratory. One of the goals of the STAR physics program is to study the internal 
spin structure of the proton in polarized proton-proton collisions.
In particular we aim to determine the gluon polarization in the proton and the flavor 
decomposition of the quark helicity densities in the nucleon sea.

STAR is capable of tracking charged particles and measuring their momenta in a high 
multiplicity environment.
The experimental setup\cite{nim} provides tracking, particle identification and  electromagnetic 
calorimetry covering a large acceptance. The identification and measurement of jets, electrons, 
photons, and neutral pions are of particular importance to the spin program.

During the first three polarized proton running periods 
at RHIC crucial machine components, including spin rotators and polarimeters\cite{nim}, 
were successfully commissioned.
Progress has been made towards the projected design luminosities and polarizations,  
$\cal{L}_{\rm{max}}$$=0.8(2.0)\times10^{32}\rm{cm^{-2}s^{-1}}$ and P=$0.7$ at $\sqrt{s}=200 (500)$ 
GeV. Specifically, polarization development has resulted in an increase in 
beam polarization from about $P= 0.15$ in the first running period in 
2002 to $P = 0.40$ in 2004, and peak luminosities have reached $0.5 
\times 10^{31}\rm{cm^{-2}s^{-1}}$.

STAR has measured the single transverse spin cross-section asymmetry $A_N=$\\
$(\sigma_{\downarrow}-\sigma_{\uparrow})/(\sigma_{\downarrow}+\sigma_{\uparrow})$ at $\sqrt{s}=200$ 
GeV in neutral pion and 
charged particle production in the forward region during the initial beam periods at 
lower luminosities and polarizations. In addition, STAR has developed a sensitive local polarimeter 
to measure the radial and transverse polarization components at the STAR Interaction Region (IR).  
We discuss the instrumentation and results below. 

A prototype of the Forward Pion Detector (FPD) was installed in STAR during 
the first pp run at RHIC. Its main component was a Pb-scintillator sampling 
calorimeter with its associated shower maximum detector for $\pi^0$ 
identification placed about $30$ cm left of the polarized beam direction
and $750$ cm from the interaction point.
Data were collected when the energy deposited in the calorimeter exceeded $15$ GeV. 
Pions were reconstructed for total energies of $15 < E < 80$ GeV with a mass resolution of $20$ MeV. 
\begin{figure}
\begin{center}
\epsfig{file=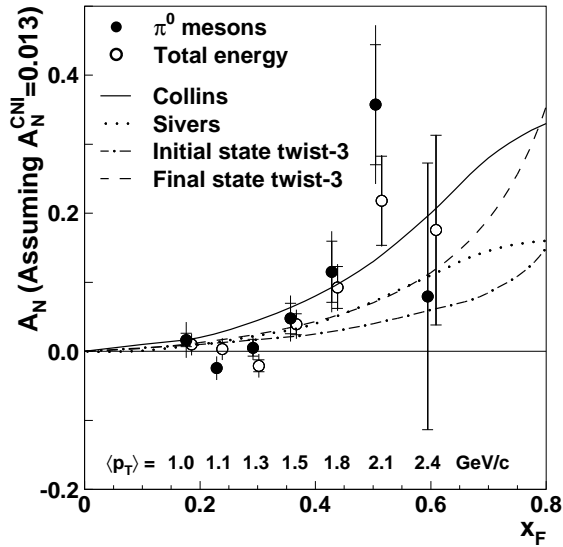,width=\pictwidth}
\caption{The analyzing power for 
$p^{\uparrow}+p\rightarrow \pi^0 + X$ at $\sqrt{s}=200$ GeV as a function of Feynman-$x$. 
The curves represent model predictions at $p_T=1.5$ GeV/c, cf. Ref.$^2$.}
\end{center}
\label{fig:fpd}
\end{figure}
Figure~\ref{fig:fpd} shows the published\cite{fpd} analyzing power for inclusive 
neutral pions production at $\sqrt{s}=200$ GeV as a function of Feynman-$x$, where 
$x_F\sim 2E/\sqrt{s}$.
The filled points are for reconstructed $\pi^0$  mesons. 
The curves are predictions from different $A_N(x_F,p_T=1.5 {\rm{GeV/c}})$ models. 
The model predictions\cite{fpd} are in qualitative agreement with the measured analyzing power, 
which is found to increase with $x_F$. Higher precision measurements of $A_N$ will map 
the dependencies in $x_F$ and $p_T$ and, together with complementary measurements, 
may help to unravel the physics mechanism.
STAR aims to further study $A_N$ for forward $\pi^0$ and aims to measure the transverse 
spin dependence of di-jet back-to-back correlations related to the Sivers function\cite{boer}. 

\begin{figure}[ht]
\begin{center}
\hspace*{-0.2in}\epsfig{file=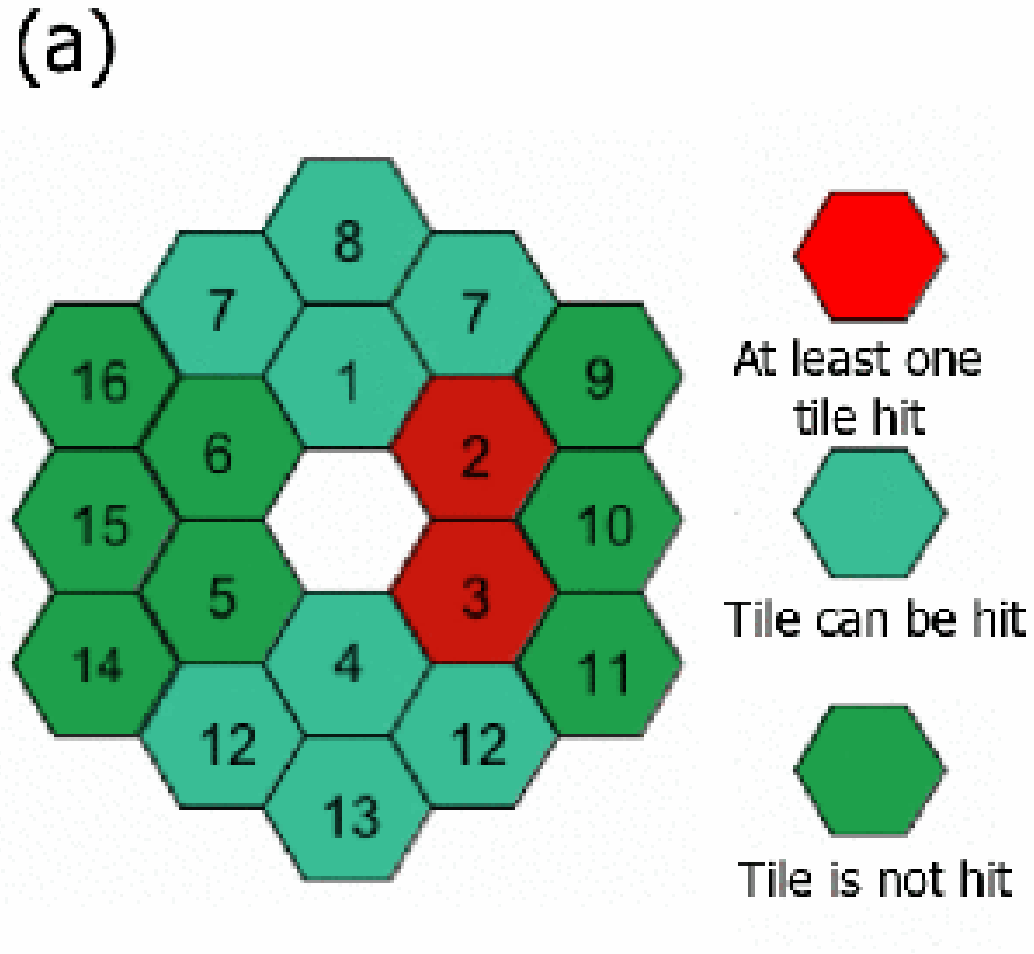,width=2.7in}\hspace*{0.1in}\epsfig{file=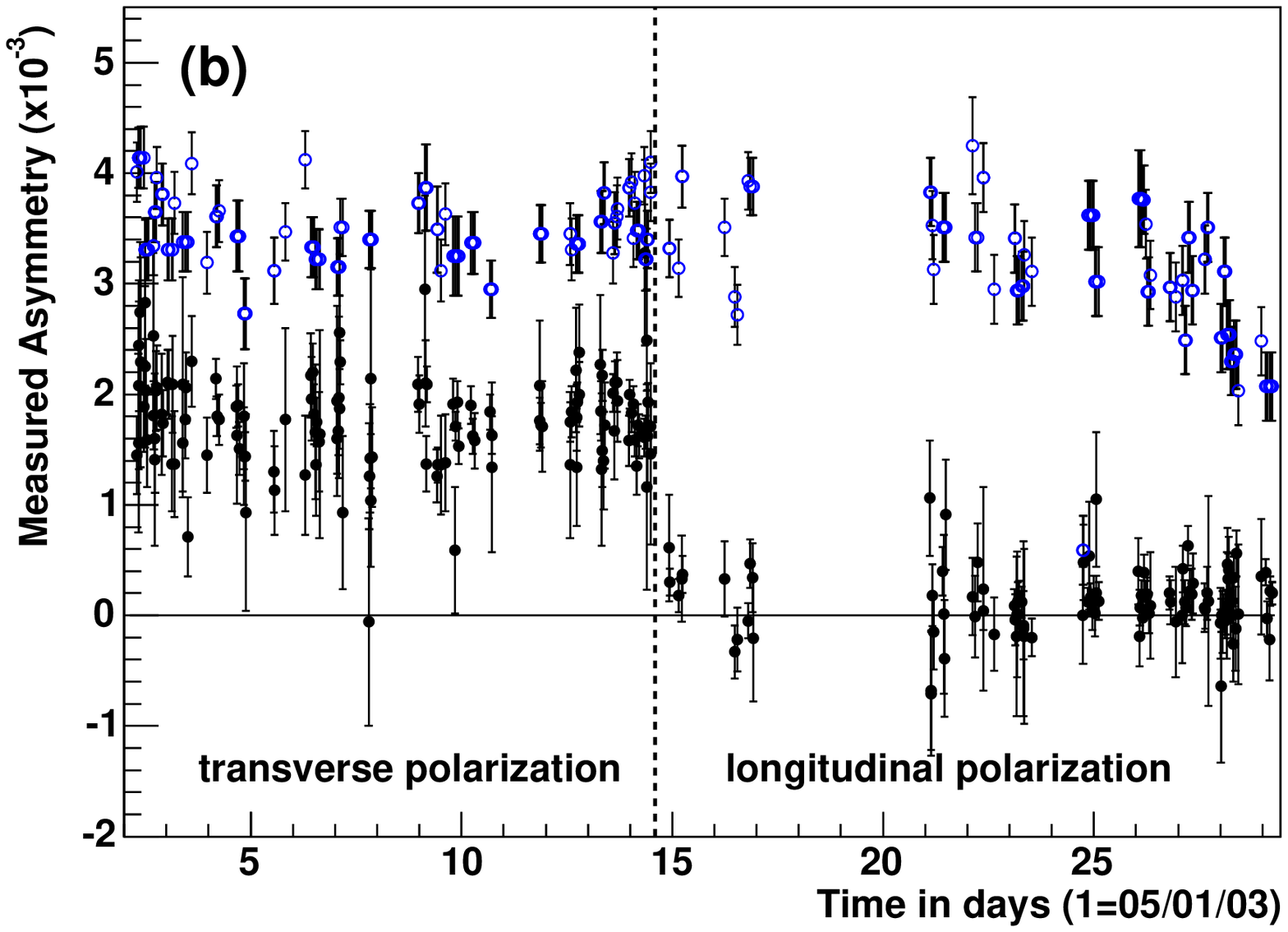,width=3.9in}
\caption{ (a) Schematic view of the BBC and event topology for charged particle {\it{Right}} 
scattering at $3.9 < \eta < 5.0$ (b) measured 'Left-Right' asymmetries (in parts per thousand)
as a function of time (in days) since May 1 2003:
BBC asymmetries (filled points) and CNI asymmetries (open points).}
\end{center}
\label{fig:bbc}
\end{figure}

The  Beam-Beam Counters (BBC) are scintillator annuli mounted around the beam pipe beyond 
the east and west poletips of the STAR magnet at 370\,cm from the interaction 
region. The small tiles of the BBC have full azimuthal coverage in the pseudorapidity range of 
$3.4 < |\eta| < 5.0$, cf. Fig.~\ref{fig:bbc}(a). 
A signal from any of the 18 tiles on the east side and any of the 18 tiles on the
west side of the interaction region constitutes a BBC coincidence.
The number of BBC coincidences is a measure of the luminosity, with
the BBC acceptance covering about 50\% of the total proton-proton cross section.
The BBC are used in STAR during proton runs as a trigger detector, to monitor the overall luminosity, 
and to measure the relative luminosities for different proton spin orientations\cite{kiryluk} 
with an accuracy of $10^{-3}$.

In an experiment with a transversely polarized beam and a left-right symmetric detector,
such as the BBC, the single spin asymmetry can be determined by measuring the beam polarization 
and the asymmetry of yields,
$$
\epsilon_{\rm{BBC}} = \frac{ \sqrt{ N_L^{\uparrow} N_R^{\downarrow} }
- \sqrt{ N_L^{\downarrow} N_R^{\uparrow} }}{
\sqrt{ N_L^{\uparrow} N_R^{\downarrow} } + \sqrt{ N_L^{\downarrow} N_R^{\uparrow} }} 
\simeq A_N^{\rm{BBC}} \times P,
$$
\noindent
in which $N_{L}^{i}$ and $N_{R}^{i}$ are the spin dependent yields
from the detector on the left ($N_L$) and right ($N_R$) side of the beam,
and $i = \uparrow, \downarrow$ denote the different spin orientations of the polarized beam.
The beam polarization $P$ was measured by the Coulomb-Nuclear Interference (CNI) polarimeter 
at RHIC\cite{cni}.  
STAR has no tracking and particle identification in  the BBC acceptance, however, 
the BBC segmentation 
allows the classification of the counted occurrences by pseudorapidity 
(2 bins: inner $3.9 < \eta < 5.0$ and outer $3.4 < \eta < 3.9$ BBC rings) and azimuth.  
The group of the 4 small tiles labeled 1, 7 and 8 in Fig.~\ref{fig:bbc}(a) 
is referred to as {\it{Up}},
whereas the group of tiles 4, 12 and 13 is called {\it{Down}}.
The remaining small tiles are labeled {\it{Left (Right) }} for the groups of tiles on 
the left (right).
An example of the hit topology for the `inner-right' BBC event is shown in Fig.~\ref{fig:bbc}(a).

Figure~\ref{fig:bbc}(b) shows the time variation of the charged particle asymmetries determined with 
the BBC for $3.9 < \eta < 5.0$ (filled points) and the asymmetry measured with the RHIC 
CNI polarimeter (open points).
Each data point corresponds to one STAR run, which typically lasts for 30-60 min. 
The indicated uncertainties on the CNI and BBC asymmetries are statistical only.
The dashed line indicates when the spin rotators at STAR were turned on and the 
transverse polarization direction in RHIC was made longitudinal at the STAR interaction region. 

From the data with transverse beam polarization at STAR we find that 
$A_{\rm{N}}^{\rm{BBC}} = 0.67(8)\times A_{\rm{N}}^{\rm{CNI}} \sim 1 \% $ for $ 3.9 < \eta < 5.0 $,
while for smaller pseudorapidities, $ 3.4 < \eta < 3.9 $, the BBC asymmetries are found to be 
$A_{\rm{N}}^{\rm{BBC}} = 0.02(9)\times A_{\rm{N}}^{\rm{CNI}}$ 
consistent with zero. 
The {\it{Left-Right}} asymmetries in the BBC are sensitive to the transverse polarization vector.  
Their numerical values when the rotator magnets were on, that is, when the beam polarization 
was longitudinal 
at the STAR IR, were significantly smaller. The currents in the rotator magnets were adjusted 
to make these 
asymmetries consistent with zero, while at the same time the CNI polarimeter - located at 
a different IR at RHIC 
- continued to measure non-zero beam polarization. The asymmetries have also been evaluated 
with the {\it{Up}} 
and {\it{Down}} groups of tiles in the BBC.  They were found to be close to zero for both 
transverse and longitudinal 
beam polarizations, as expected. 

\begin{figure}
\begin{center}
\epsfig{file=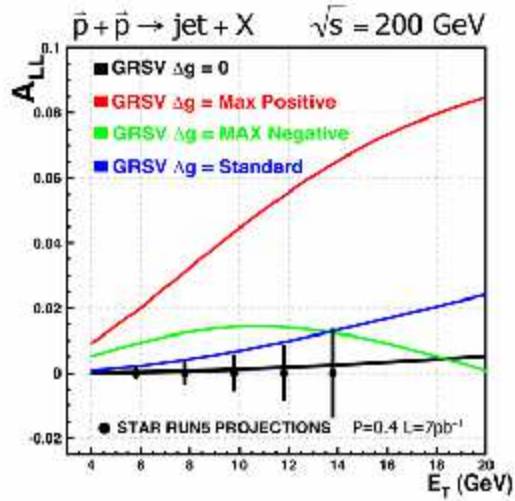,width=\pictwidth}
\caption{ Projected statistical uncertainties on the double longitudinal spin asymmetry $A_{LL}$ 
in inclusive jet production for 40\% longitudinal beam polarizations (as achieved in 2004) and 
an integrated luminosity of 7\,$pb^{-1}$ (recorded). The curves represent theoretical predictions 
using polarized parton distribution functions from Ref.$^{10}$ with various assumption on the gluon 
polarization at the initial scale.}
\end{center}
\label{fig:jets}
\end{figure}

STAR has thus far collected about $1$ pb$^{-1}$ of data at $\sqrt{s}=200$ GeV with longitudinally 
polarized beams.
These data will allow an exploratory measurement of the double longitudinal spin asymmetry $A_{LL}$ 
in inclusive 
pion and jet production, which is sensitive to the magnitude of the gluon polarization in 
the proton\cite{jager}.
At the partonic level, the cross section receives contributions from the (a) $g+g\rightarrow g+g$, 
(b) $g+q \rightarrow g+q$ and (c) $q+q \rightarrow q+q$ processes.
Their relative contribution varies with $p_T$ and is dominated by gluon-gluon and quark-gluon 
scattering at low $p_T$.
Preliminary precisions and the status of the ongoing jet analysis of the existing data 
have been reported 
in Ref~\cite{trentalange}. The ongoing completion of the Barrel Electromagnetic Calorimeter, 
and the successful 
installation of an Endcap Electromagnetic Calorimeter, will expand STAR's acceptance and 
triggering capabilities 
for pions and jets.  Figure~\ref{fig:jets} shows prospects for the upcoming running period in 2005. 

In the longer term, when the STAR calorimeter upgrades have been completed and design polarizations 
and beam luminosities have been reached, we will measure the double longitudinal spin asymmetry 
$A_{LL}$ for coincident photon jet production $\vec{p}+\vec{p}\rightarrow \gamma + {\rm{jet + X}}$ 
at both 200 and 500 GeV 
center-of-mass energy to determine the gluon polarization over a wide range in $x$.  
At leading order QCD, the prompt photon production in $pp$ collisions is dominated
by the gluon Compton process $q+g \rightarrow \gamma+q$ and $A_{LL}$ can be written as:
$$
A_{LL} \simeq {\frac{\Delta G (x_g,Q^2)}{G(x_g,Q^2)}} \times A_1^p (x_q,Q^2) 
\times \hat{a}_{LL}^{\rm Compton}.
$$
The proton asymmetry $A_1^p$ is known from inclusive DIS measurements
and the partonic asymmetry for the Compton process
$\hat{a}_{LL}^{\rm Compton}$ can be calculated in perturbative QCD.
The QCD scale $Q^2$ is on the order of the $p_T^2$ of the prompt photon,
and the fraction $x_q (x_g)$ of the hadron momentum carried by the quark
(gluon) can be reconstructed on an event-by-event basis
from the observed photon and jet pseudorapidities and the transverse momentum 
of the photon\cite{bland}.
The measurement of $A_{LL}$ thus forms a direct determination of
$\Delta G(x,Q^2)/G(x,Q^2)$. 
Theoretical description of $A_{LL}$ is also known at next-to-leading order~\cite{frixione}.
The measurement for two values of $\sqrt{s}$ is essential to cover
a relatively wide region in $x_g$, $0.01 < x_g < 0.3$.
The integral $\Delta G = \int_{0}^{1} \Delta G(x)$ d$x$ from these
measurements is expected to be determined to a precision better than $\pm 0.5$ 
assuming  integrated luminosities of \mbox{320\,pb$^{-1}$}
at \mbox{$\sqrt{s} = 200\,\mathrm{GeV}$} and
\mbox{800\,pb$^{-1}$} at \mbox{$\sqrt{s} = 500\,\mathrm{GeV}$}~\cite{bland}.
The upcoming proton run (2005-2006) will be the first run for which all of the STAR 
detector components essential for this measurement are commissioned. 

In addition to the measurements of the gluon polarization,
STAR aims to  decompose the quark spin densities in the nucleon sea by
measuring the parity violating single spin asymmetries for $W$
production in $\vec{p} + p \rightarrow W + X \rightarrow e + X$
collisions at $\sqrt{s} = 500\,\mathrm{GeV}$.
At these energies, $W$ bosons  are produced predominantly through 
$u + \bar{d} \rightarrow W^+ $ and 
$d + \bar{u} \rightarrow W^-$, valence-sea processes  
in  $pp$  collisions.  
The measurements, which will require the operation of RHIC at high luminosities and 
an upgrade of the tracking capability in the forward region at STAR, 
are expected to aid the flavor decomposition of the quark spin 
densities in the nucleon sea\cite{soffer} and to distinguish various symmetry scenarios.  
Their theoretical description is known at next-to-leading order~\cite{nadolsky}.

\end{document}